\documentclass[twocolumn,amsmath,amssymb,longbibliography,superscriptaddress,showkeys]{revtex4-1}
\pdfoutput=1

\usepackage{graphicx}
\usepackage{dcolumn}
\usepackage{bm}
\usepackage{algpseudocode} 
\usepackage{subfigure} 
\usepackage[braket]{qcircuit}  

\usepackage{color}
\usepackage{url} 
\usepackage[normalem]{ulem} 
\usepackage{xcolor}
\usepackage{epsfig} 
\usepackage[colorlinks]{hyperref}  
\usepackage{amsmath}
\usepackage{amssymb}
\usepackage{float}
\usepackage{mathtools}
\usepackage{braket} 

\newcommand{\re}{{\rm I\!R}}

\begin{document}  

\title{Robust Implementation of Generative Modeling with Parametrized Quantum Circuits}%
 
\author{\mbox{Vicente Leyton-Ortega}}
\affiliation{Rigetti Computing, 2919 Seventh Street, Berkeley, CA 94710-2704, USA}
 
\author{Alejandro Perdomo-Ortiz}
\email{Correspondence: ucacper@ucl.ac.uk}
\affiliation{Rigetti Computing, 2919 Seventh Street, Berkeley, CA 94710-2704, USA}
\affiliation{Department of Computer Science, University College London, WC1E 6BT London, UK}

\author{Oscar Perdomo}
\affiliation{Rigetti Computing, 2919 Seventh Street, Berkeley, CA 94710-2704, USA}
\affiliation{Department of Mathematics, Central Connecticut State University, New Britain, CT 06050, USA}

\begin{abstract}

    Although the performance of hybrid
    quantum-classical algorithms is highly dependent on the
    selection of the classical optimizer and the circuit
    ansatz~\cite{Benedetti2018,Hamilton2018, Zhu2018, Perdomo2019a}, a robust
    and thorough assessment \textit{on-hardware} of such features has been
    missing to date. From the optimizer perspective, the primary challenge lies
    in the solver's stochastic nature, and their significant variance over
    the random initialization. Therefore, a robust comparison requires that one
    perform several training curves for each solver before one can reach
    conclusions about their typical performance. Since each of the training
    curves requires the execution of thousands of quantum circuits in the
    quantum computer, such a robust study remained a steep challenge for most
    hybrid platforms available today. Here, we leverage on Rigetti's Quantum
    Cloud Services (QCS\texttrademark) to overcome this implementation barrier, and we study
    the \textit{on-hardware} performance of the data-driven quantum circuit
    learning (DDQCL) for three different state-of-the-art classical solvers, and
    on two-different circuit ans\"{a}tze associated to different entangling
    connectivity graphs for the same task. Additionally, we assess the gains in
    performance from varying circuit depths. To evaluate the typical performance
    associated with each of these settings in this benchmark study, we use at
    least five independent runs of DDQCL towards the generation of
    quantum generative models capable of capturing the patterns of the canonical
    Bars and Stripes data set. 

\end{abstract}


\maketitle

\section{\label{sec:Intro} Introduction}

With the advent of several quantum computing technologies available to date, a
significant effort is devoted to finding algorithmic strategies to cope with the
noise is such early hardware architectures. In this domain, hybrid
quantum-classical (HQC) algorithms, such as the variational quantum eigensolver
(VQE)~\cite{Peruzzo2014,McLean2016} and the quantum approximate optimization
algorithm (QAOA)~\cite{Farhi2014}, provide ways to use quantum
computing hardware for practical applications. On the other hand,
characterization of these so-called noisy-intermediate scale quantum (NISQ)  
devices is also one of the significant endeavors, since it paves the way to
realizing the power of these devices. Covering both of these aspects, the
data-driven quantum circuit learning (DDQCL)\cite{Benedetti2018} algorithm was
proposed to provide not only a framework to probe the power of NISQ devices on a
useful machine learning setting (i.e., the case of generative modeling in
unsupervised machine learning) but also providing a way to measure the power of
HQC variants. Some of these might include but not be limited to the choice of the
circuit ansatz where the entangling connectivity layout and type of gates need
to be specified, as well as the selection among different optimizers to be used
in the classical processing side.

\begin{figure}  
  \centering 
    \includegraphics[width=0.4\textwidth]{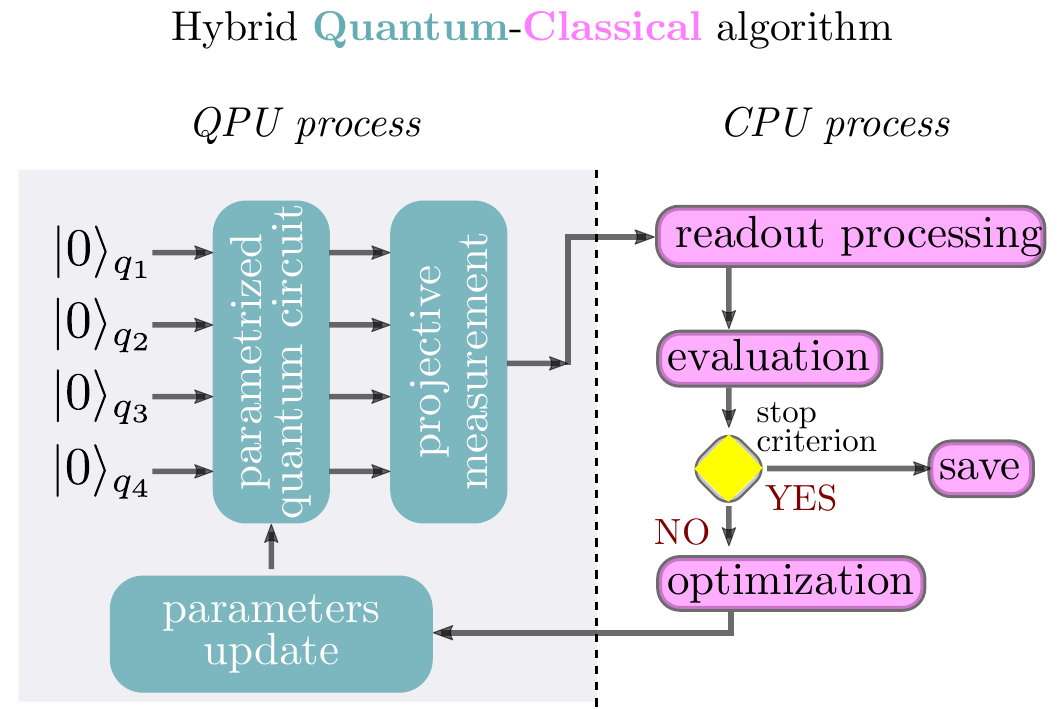}   
    \caption{{\it Hybrid quantum-classical algorithm for data-driven quantum
    circuit learning (DDQCL)}. Sketch for a
    simple case of four qubit setup, in which, the probability distribution of the
    resulting quantum state is used as the generative model for the ML task. Training
    of the quantum circuit towards the generation of such quantum model is achieved by a continuous update of the quantum
    circuit parameters by a classical optimizer. The quantum model distribution is obtained from simple projective measurements in the computational basis, which, after readout processing, it is compared with the
    target distribution dictated by the data set. The optimizer updates the quantum circuit
    parameters until a stop criterion is met, e.g., a maximum number of
    iterations or a desired value of the optimization cost function is reached.           
      \label{fig:algo}}  
\end{figure}

Although used for different tasks, DDQCL and other variational algorithms
approaches have in common components that need to be fine-tuned
towards a successful experimental implementation (see Fig.~\ref{fig:algo}). The
quantum processing side is represented by a parametrized quantum circuit (PQC)
whose parameters are updated via an optimizer running on a classical processing
unit. Fine tuning of both of these components is essential to the performance of
the overall HQC implementation.

Experimental implementations of DDQCL where the entire learning process
performed \textit{on-hardware} have been recently demonstrated in both,
superconducting qubits~\cite{Hamilton2018} and ion-trap quantum
computers~\cite{Zhu2018}. In the superconducting qubits implementation emphasis
was given to the exploration of different circuit ans\"{a}tze, while in the
ion-trap experiments, in addition to consideration of different proposals for
the quantum processing unit (QPU) side of the pipeline, emphasis was given as well to the importance of
the optimizer used.  

Each of the independent training curves of DDQCL and any HQC algorithm requires
several iterations which correspond to the updates of the PQC. This procedure,
in turn, represents typically thousands of circuits that need to be executed in
hardware, per training curve. As seen in any of the aforementioned experimental implementation of DDQCL
\textit{on-hardware}, usually only one learning
curve is performed, for each of the components to be optimized, e.g., circuit ansatz or
optimizer. As shown in this work and elsewhere~\cite{Benedetti2018}, for most
of the optimizers previously considered in the literature, the variance from the
solver's random initialization is one of the most significant sources of
uncertainty when reporting the performance of the HQC algorithm. Thus, one
successful run does not necessarily correlates with the typical performance of
the solver; none of the studies to date seem to afford multiple runs per
each of the settings studied. To assess the impact of any choice within the HQC algorithm
and to draw robust conclusions about the impact of several entangling
connectivities or the performance of any optimizer against another, several runs per
setting need to be performed. 

In this work we present the first experimentally robust comparison of each of
the aspects of HQC algorithms, using DDQCL as the working framework since it
provides figure merit for the performance of each of the knobs explored.  To achieve this systematic and robust study, we leverage on Rigetti's Quantum Cloud
Services (QCS\texttrademark) and its enhanced framework for handling the requirements of HQC
algorithms efficiently. Some unique features of this platform include pre-compilation of the PQC programming cycles, active reset allowing for faster repetition rates, and low-latency due to co-location of the quantum and classical hosts.

\section{DDQCL execution on Rigetti's QPU}

\begin{figure}[H]  
  \centering 
    \includegraphics[width=0.3\textwidth]{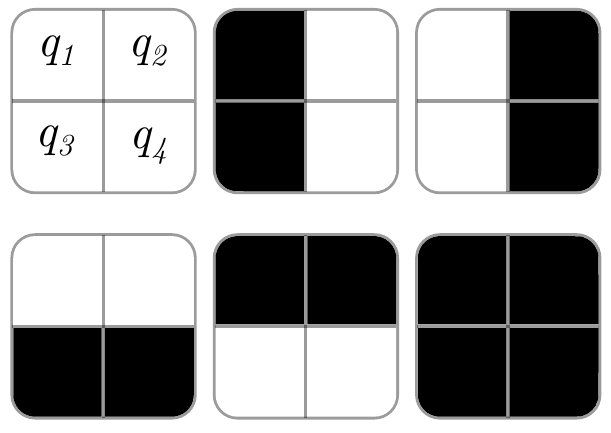}    
    \caption{{\it $(2,2)$ BAS classical data set}. The patterns that belong to
      $(2\times 2)$ Bars and Stripes dataset are shown. In the first pattern (on
      the top left corner), the label convention is depicted to encode the
      patterns into classical bitstrings, e.g., for the second binary image
      (from the top left corner to the right), $(q_1,q_2,q_3,q_4) =
      (\blacksquare, \Box,  \blacksquare, \Box ) = (1,0,1,0) \rightarrow 1010$.
    \label{fig:bas}}  
\end{figure}

As in previous \textit{on-hardware} implementations of
DDQCL~\cite{Hamilton2018,Zhu2018}, for the machine learning task we consider a
small artificial classical dataset, Bars and Stripes (BAS), which consists of
patterns of $n \times m$ binary images of bars or stripes, and denoted here as
$(n,m)$BAS. As shown in Fig.~\ref{fig:bas}, the $(2,2)$BAS dataset contains six
different binary images. We encode the binary imagines into bitstrings,
associating dark pixels to ``1" and white pixels to ``0", i.e. $\blacksquare =
1$ and $\Box = 0$. In Figure \ref{fig:bas}, we present the patterns for
$(2,2)$BAS and the label convention to use for binary image encoding. Following
that convention the $(2,2)$BAS dataset contains $\{0000, 1010, 0101, 0011, 1100,
1111 \}$ that leads to a simple distribution $P_{\cal X}(\bm{x}) = 1/6$ for
$\bm{x} \in$ $(2,2)$BAS. The goal for the circuit learning approach is to
prepare a state $|\psi\rangle = \sum_{\bm{x} \in \{0,1\}^4 } \alpha_{\bm{x}} |
\bm{x} \rangle $ with $|\alpha_{\bm{x}}|^2 = 1/6$ for $\bm{x} \in (2,2)$BAS and
$|\alpha_{\bm{x}}|^2 = 0$ otherwise.
 For the experimental implementation of DDQCL, we consider a four-qubit subunit
 connected in {\it star} and {\it line} entangling topology, according to the
 connectivity constraints in the 16-qubit Aspen\texttrademark quantum computer  (see insets in
 Fig~\ref{fig:main}). The main building blocks for the construction of the quantum circuit
 ansatz (see, e.g. top right in Fig.~\ref{fig:main}) relies on two observations. (i) The use
 of  CZ entangling gates is native to the Rigetti QPU. This sets our preference
 towards these gates over either CNOTs or M{\o}lmer-S{\o}rensen gates which were
 used in previous DDQCL implementations \cite{Hamilton2018,Zhu2018} on either
 the IBM Tokyo or ion-trap quantum computer, respectively. (ii) Instead of using
 arbitrary single-qubit gates intersperse or alternating with two-qubits gates,
 we used here just $R_y(\theta)$ rotations. Given that all the entries of the
 $R_y(\theta)$ are real, along with the CZ gates produces quantum states with
 real amplitudes only. Note this is a not limitation towards the quantum model
 used in the generative task since it is enough to consider
 $\alpha_{\bm{x}} \in \re$. In Appendix~\ref{ss:ansatz}, we provide more intuition of the
 layout for this circuit ansatz. This last choice of going from single-arbitrary
 rotations to just $R_y$ rotations is desirable since it reduces the number of
 parameter in the quantum model, therefore helping the classical optimizer in
 this high-dimensional search space.

The progress of this machine learning task is evaluated by the Jensen-Shannon
divergence cost function (details in Appendix~\ref{s:ddqcl}), that compares the probabilistic
distribution of the circuit output with the target distribution. With the
learning score record, the quantum circuit parameters are updated by an
optimizer running on a classical processor; this optimization in pursuit of a
minimum score is the classical part of the generative model approach.

\section{results} \label{s:results}

For our experimental and numerical benchmarking study we used three solvers: the
Zeroth-Order Optimization package (ZOOPT)~\cite{Liu2017}, the Stochastic
Variation of Hill-Climbing type algorithm (SVHC)~\cite{Perdomo2019a}, and a
classical stochastic gradient descent based solver (ADAM)~\cite{Kingma2014} (for
configuration details, see Appendix~\ref{ss:optimizers}).

\begin{figure*}
 \includegraphics[width=0.80\textwidth]{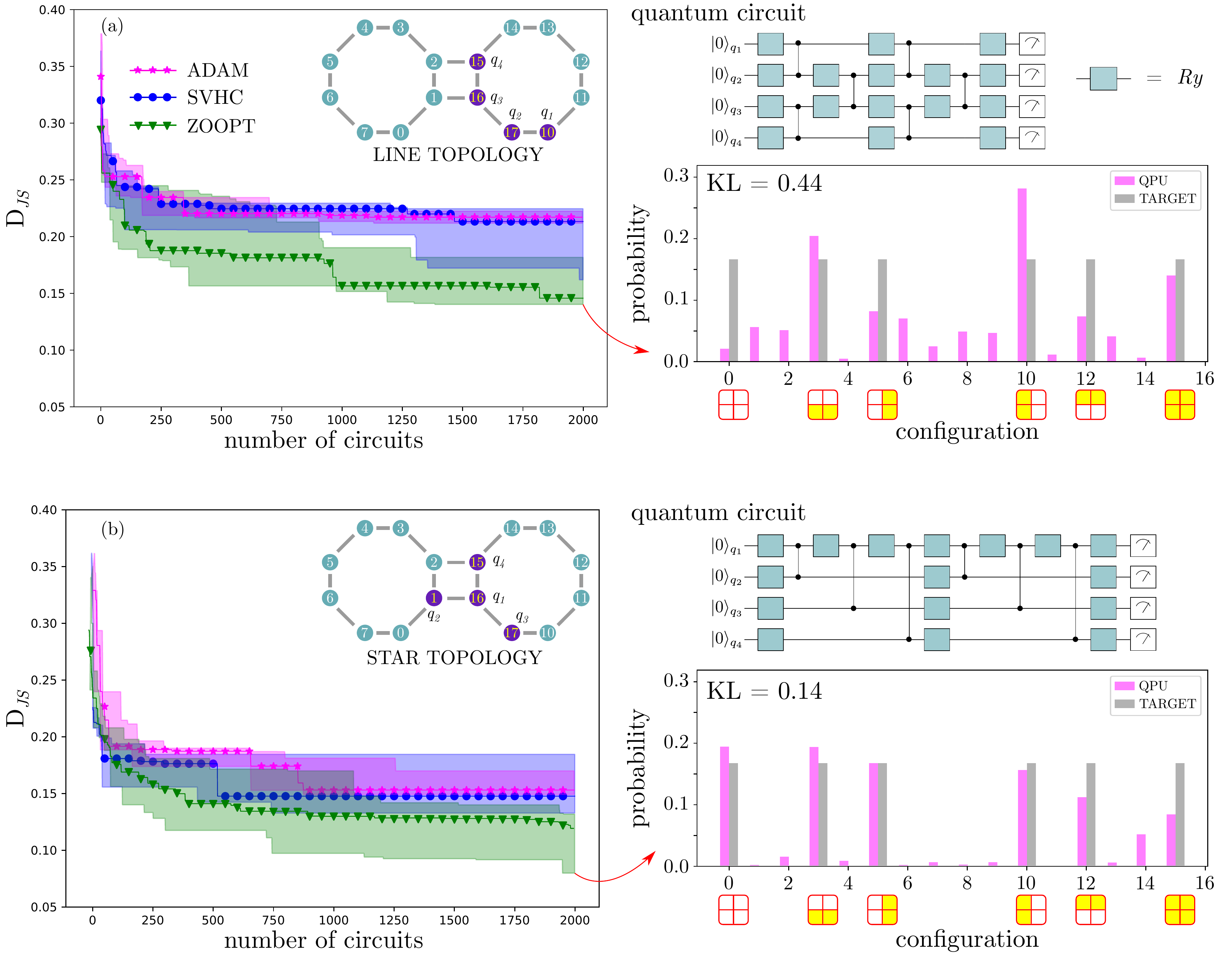}    
 \caption{{\it DDQCL on the $(2,2)$BAS and Stripes dataset, using different
   solvers and topologies}. In the upper/lower panel (a/b) is depicted the
   learning progress for two layers of entangling circuits with a
   \textit{line/star} topology connectivity using ADAM, SVHC, and ZOOPT
   classical optimizers. Besides, in the panels are shown the quantum circuit
   model and the comparison between the probability distribution corresponding
   to the lowest cost and the target distribution. In that comparison, it is
   shown the Kullback-Leibler divergence cost to get to the target distribution
   from the  QPU best result.}    
\label{fig:main} 
\end{figure*}

We consider a specific region of the QPU to run the experiments (see QPU layout
inside Figure~ \ref{fig:main}). For each of the settings to be explored,  i.e.
the choice of solver or the entangling connectivity topology or the varying
depth of the circuit, we performed five independent DDQCL from random
initializations of the parameters. In each DDQCL, 2000 different realizations
of the PQC were evaluated, and from each of them, 3000 shots were taken as
readouts in the computational basis. Before every batch of these five
independent runs, we performed the characterization of the matrix to be used for
readout correction and use it to post-process the experimental histograms built
from every 3000 shots (see Appendix~ \ref{ss:readoutcorr} for details). In
Figure~\ref{fig:main}, we present the main results, where the best learned
probabilistic model is depicted with $D_{KL}$ values 0.44 and 0.14 for {\it
line} and {\it star} topology, respectively. Those values were reached using the
ZOOPT optimizer. $D_{KL}$ corresponds to the Kullback-Leibler divergence and it
is the gold standard when comparing the closeness between two probability
distributions, with $D_{KL} = 0.0$ meaning that the quantum model and the target
distribution match. Estimation of $D_{KL}$ from the experimental histogram also
allow us to compare to previous experimental realizations~\cite{Hamilton2018,
Zhu2018}. From these it can be seeing that our DDQCL implementation on the
Rigetti QPU ($D_{KL} = 0.14$) is better  than the best model obtained in the IBM
Tokyo quantum processor ($D_{KL} = 0.36$), and comparable to the best results
obtained in the trapped-ion quantum computer ($D_{KL} = 0.09$) 

\begin{figure*} 
\includegraphics[width=0.7\textwidth]{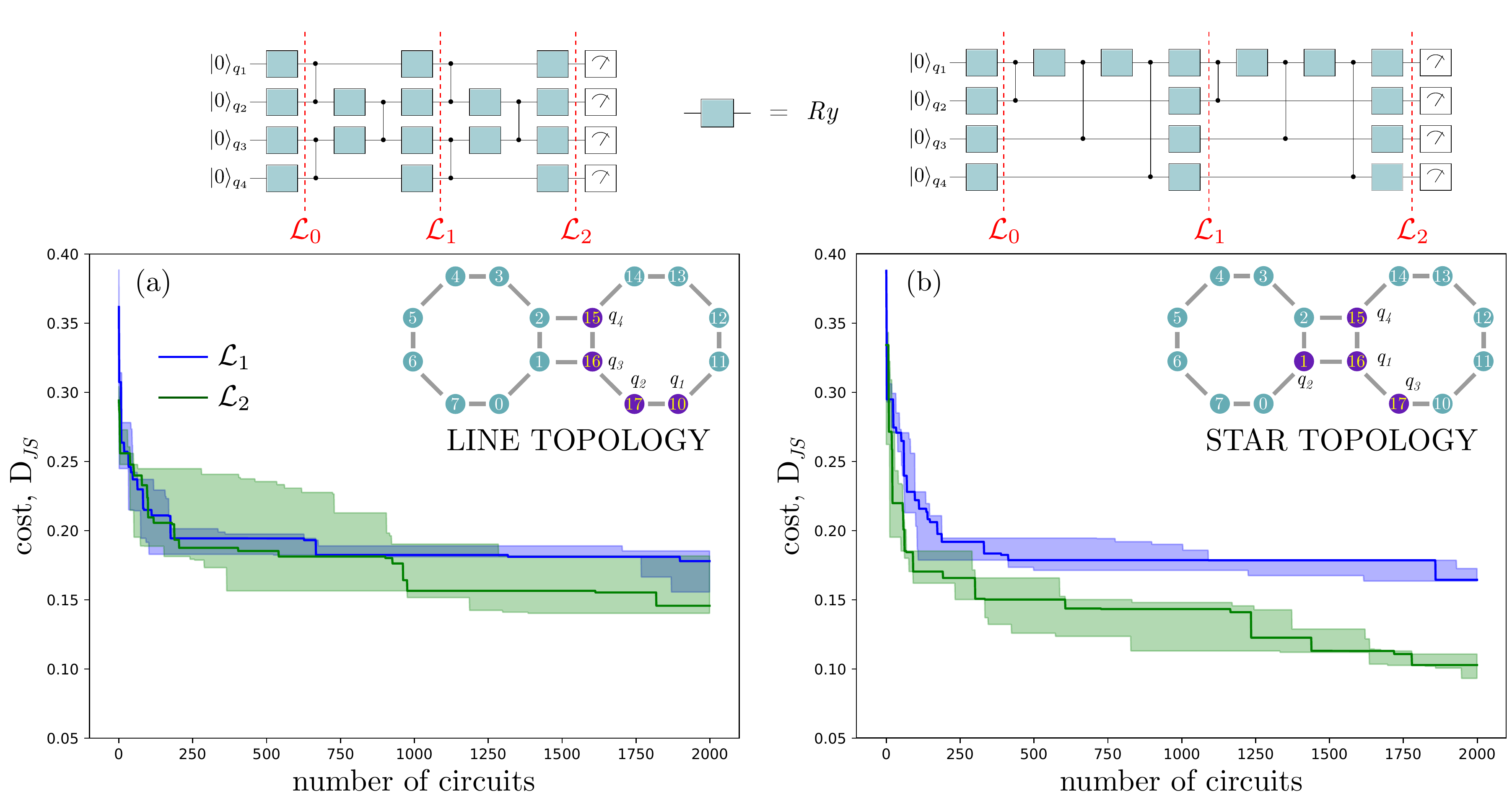}
\caption{{\it Learning improvement with circuit depth}. In this Figure, we show
  the learning improvement with the circuit depth on the {\it line} and {\it
  star} topology. We define an entangling layer as a set of entangling gates ${\cal
  U}_2$ overall qubits according to the topology. For line topology, after
  simplification, a layer for 4 qubits consists of 6 rotations $R_y$ and 3 CZs,
  see the quantum circuit over panel (a). The same amount of rotations $R_y$ and
  CZ's stands for the {\it star} topology, see the quantum circuit over panel
  (b). A relative improvement due to circuit depth is presented using star
  topology against the line topology, in which the variance of the learning
  curve for 1 layer overlaps the variance of the learning curve for 2 layers. In
  the experiments we consider qubits in similar regions of the device (see
  layouts inside) according to the topologies. }
\label{fig:layers}   
\end{figure*}

In addition to testing the performance of different types of optimizers, we
consider different circuits depths. For these experiments, we consider a circuit
with one round of entangling gates with  3 CZ's and 10 local rotations, and the
circuit used for the optimizers comparison corresponding to two rounds of
entangling gates with 16 local rotations and 6 CZ's (see ${\cal L}_1$ and ${\cal
L}_2$ in Figure~\ref{fig:layers}, respectively). In {\it star} topology
configuration, a significant improvement is observed with the addition of a
second entangling round, while for the {\it line} topology configuration, there
is significantly less improvement (or at least slower convergence) from the
addition of a second entangling round. To understand this apparent faster
convergence and better results from the star entangling topology compared to its
line analog, we performed \textit{in-silico} simulations with the same
optimizers. From these simulations on noiseless qubits (see
Appendix~\ref{s:simulations}), it is hard to advocate for a clear advantage of
the star connectivity over the line entangling ansatz. Therefore, this
discrepancy in results between {\it in-silico} simulations and the on-hardware
ones is not to the topology itself, but most likely it is due to the quality of
the qubits used for each experiment, with better quality for the subset used for
the star experiments. 

For completeness, in Table~\ref{t:experiments} we report the values for the qBAS
score, calculated as detailed in Ref.~\cite{Benedetti2018}.

\begin{table}[h] 
\caption{Performance figures of merit from main experimental
features explored in this work related to different optimizers and circuit
ans\"{a}tze (star versus line).}
\label{t:experiments} 
\includegraphics[width=0.45\textwidth]{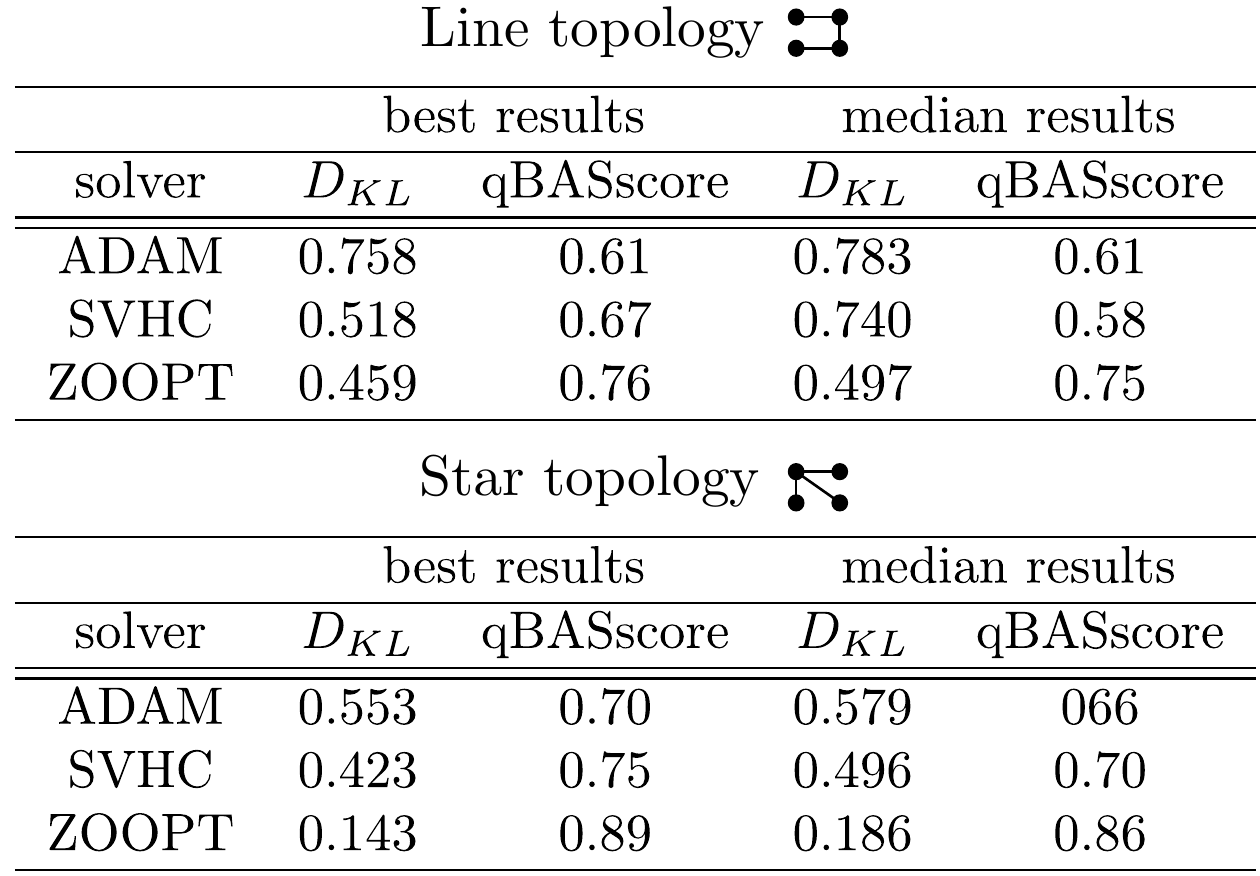}
\end{table}

\section{Summary and outlook}

We performed a robust implementation and a comparison {\it on-hardware} of several
components of primal importance affecting the performance of hybrid
quantum-classical algorithms. The factors tested include the circuit
entangling layout and its depth, the selection of the classical optimizer, and
the impact of post-processing strategies such as readout correction. 

Although results here show that the choice of each of these components affects
the performance significantly, the study presented here is far from exhaustive.
It is known that optimizing each of the classical solvers in its own is a hard
task, and it would be left to future work to address this selection, for
example, by using an automatic routine for hyperparameter setting (see, e.g.,
\cite{Snoek2012}).

Another interesting direction would be to try some recent extensions of DDQCL
experimentally, proposing a gradient-based training
approach~\cite{liu2018differentiable} or by adding ancillary qubits to enhance
the power of the quantum generative  model~\cite{du2018expressive}.

We hope hybrid approaches on concrete data sets and real-world applications,
such as the generative modeling task highlighted here, become a more significant
way to benchmark and measure the power of quantum devices. The qBAS score
described in Ref.~\cite{Benedetti2018} and measured here for our experimental
results represents a way to measure the power of these NISQ devices, beyond
values reported for the fidelity of single and two-qubit gates. To move to a
quantum ready stage, it is essential we move towards evaluating the capacity and
performance of the device as a whole. We believe developing the infrastructure
to develop robust benchmarking studies like the one presented here is an
essential step towards a meaningful comparison among different algorithmic and
hardware proposals.

\acknowledgements  

The authors would like to thank Marcus P. da Silva for useful feedback and
suggesting the readout correction used in our experiments here, Marcello
Benedetti for general discussions on generative modeling and their evaluation,
and the Rigetti's QCS team for access and support during the beta testing of
the Aspen processor.

\pagebreak
\widetext

%

\appendix

\section{DDQCL pipeline}~\label{s:ddqcl}

In this work we implement a hybrid quantum-classical algorithm for unsupervised
machine learning  tasks introduced in \cite{Benedetti2018} at Rigetti's superconducting
quantum computer (QPU). In the following we present a short introduction of the
approach, for a complete  discussion see \cite{Benedetti2018}. The algorithm generates
a probability distribution model for a given classical dataset ${\cal X} =
\lbrace \bm{x}^{(1)}, \bm{x}^{(2)}, \cdots, \bm{x}^{(D)} \rbrace$ with
distribution $P_{\cal X}$.  Without lost of generality, the elements of ${\cal
X}$ can be considered as $N$-dimensional binary vectors $\bm{x}^{(i)} \in
\lbrace 0,1\rbrace^N$ for $i = 1, ...., D$, allowing a direct connection with
the basis of an $N$-qubit quantum state, i.e., $\bm{x}^{(i)} \rightarrow |
\bm{x}^{(i)} \rangle = | x_1^{(i)},...,x_N^{(i)} \rangle $ for $i = 1, ..., D$.
The goal is to prepare a quantum state $|\psi \rangle =  \sum_{\bm{x} \in \{ 0,1 \}^N } \alpha_{\bm{x}} \,
|\bm{x} \rangle $, with a probability distribution that mimics the dataset
distribution, $|\langle \bm{x} | \psi \rangle |^2 = |\alpha_{\bm{x}}|^2 = P_{\cal
X} (\bm{x})$ for $\bm{x} \in \{0,1 \}^N$.

The required quantum state $| \psi \rangle $ is prepared by tuning a PQC composed
of single rotations and entangling gates with fixed depth and gate layout. In
general, the circuit parameters of amount $L$ can be written in a vector form
${\bm \theta} = \lbrace \theta_1, ...,\theta_L\rbrace$, that prepares a state
$|\psi_{\bm \theta} \rangle$ with distribution $P_{\bm \theta}$. The algorithm
varies $\bm \theta$ to get a minimal loss from  $P_{\bm \theta}$ to $ P_{\cal X}
$.  Here, we consider the Jensen-Shannon divergence ($D_{\rm JS}$) to measure
the loss, and determine how to update $\bm \theta$ using a classical
optimizer. 

The $D_{\rm JS}$ divergence is a symmetrized and smoothed version of the
Kullback-Leibler divergence ($D_{\rm KL}$), defined as
\begin{equation}
\label{eq:JS}
	D_{\rm JS}(P|Q) = \frac{1}{2} D_{\rm KL}(P | M ) + \frac{1}{2} D_{\rm KL}(Q | M ),
\end{equation}
where  $P$ and $Q$ are distributions, and $M = (P + Q)/2$ their average. The Kullback-Leibler divergence is defined as 
\begin{equation}
  \label{eq:KL}
  D_{\rm KL}(X | M) = 
  \sum_{s \ \in \ \{ 0,1 \}^N} X(s) \,{\rm ln} \left( X(s) \right )  - 
  \sum_{s \ \in \ \{ 0,1 \}^N} X(s) \,{\rm ln} \left ( M(s) \right), 
\end{equation}
for $X \in \{ P, Q  \}$. 

After the algorithm evaluates the cost from $P_{\bm \theta}$ to $ P_{\cal X} $,
it follows a quantum circuit parameters update in pursuit to a minimal cost.
This procedure is done by an optimizer that runs on a classical processor unit
(CPU). In summary, the algorithm chooses a random set of parameters $\bm \theta$,
evaluates the cost from $P_{\bm \theta}$ to the target $P_{\cal X}$ through the
Jensen-Shannon divergence $D_{\rm JS} ( P_{\bm \theta} | P_{\cal X} )$, and
updates $\bm \theta$ minimizing the cost. This procedure is
repeated several times (iterations) until a convergence criterium is met, for
instance, a maximum number of iterations.

\section{Experimental details}

\subsection{Quantum circuit model design}\label{ss:ansatz}
Since the quantum state probabilistic distribution does not depend on local
quantum state phases, we chose a layout of the model circuit that prepares real
amplitude quantum states. We consider the following two-qubit ansatz as the base
to design $N$-qubit the circuit layout for DDQCL,  
\begin{eqnarray}
  {\cal U}_{2}(\theta,  \gamma, \beta) |00\rangle &=&
  R_1^{\beta}\, R_0^{\gamma} \, {\rm CZ}_{01}\, R_0^{\theta}\, |00 \rangle
  \nonumber \\
  &=& \sum_{s \in \{0,1\}^2} \! \!\!\alpha_s |s\rangle 
\end{eqnarray}  
therein $R_k^\theta = \exp [-i\, \theta\, \sigma^y_k /2]$ is a local rotation
around $y$-axis in the Bloch sphere on the $k$th qubit and ${\rm CZ}_{kl}$ the
control phase shift gate between $k$th and $l$th qubits.  The gate ${\cal U}_2$
defines a one-to-one map $[0, 2 \pi)^3 \leftrightarrow S^3$, that ensures the
preparation of any 2 qubit state with real probability amplitudes \cite{Perdomo2019b}, i.e. $\alpha_{s} \in \re$ for $s \in \{ 0 , 1 \}^2$. For setups with
more than 2 qubits, we use ${\cal U}_2$ to entangle different pairs of qubits
depending on their connectivity, until getting a fully controlled entangled
state. In this study, we consider four qubits with different connections
(topologies) according to the Rigetti's quantum computer architecture. We 
designed several circuits for {\it line} and {\it star} topologies using up to
$L=16$ local rotations and 6 CZ's (see Figure \ref{fig:main}). 

\textit{In-silico} simulations on noiseless qubits show that this
circuit layout can learn a quantum model capable of successfully
reproducing the target probability from the $(2,2)$BAS (see
Appendix~\ref{s:simulations}).

As pointed in the main text, one of the advantages of this circuit layout
compared to the one considering arbitrary single-qubit rotations is the
reduction in the number of parameters. On the other hand, it is important to
note that having the additional parameters could in principle help the search
since these open more paths in the Hilbert space, resulting in an increased
number of states that could lead to a perfect $P_{\cal X}$ (see e.g., the
discussion in the ``Entanglement entropy of BAS(2,2)" section in the Supp.
Material of Ref.~\cite{Benedetti2018}). This trade-off between flexibility in
the quantum model and difficulty in optimization is beyond the
scope of this work, and it would be an interesting research direction to
explore.

\subsection{Classical optimizers}\label{ss:optimizers}

For our experimental and numerical benchmarking study we used three solvers: the
Zeroth-Order Optimization package (ZOOPT) \cite{Liu2017}, the Stochastic Variation
of Hill-Climbing type algorithm (SVHC)~\cite{Perdomo2019a}, and a classical stochastic gradient
descent based solver (ADAM)~\cite{Kingma2014}.  For the case of ZOOPT and SVHC, we
handle the stochastic nature of the cost function via the "value suppression"
setting.  Additionally, a set of $N_{\rm ini} = 3L$ randomly chosen initial
circuits are evaluated to get a searching starting point. For ADAM, we consider the setup
used in Ref.~\cite{Hamilton2018}, with learning rate $\alpha = 0.2$, decay
rates $\beta_1 = 0.9$, and $\beta_2 = 0.999$.  

\subsection{Improvement via readout correction}~\label{ss:readoutcorr}

DDQCL is based on the probability distribution of the $N$ qubit quantum state,
which is in turn transformed into resulting bitstrings $\{0,1 \}^N$, denoted as
{\it shots} in the main text. Unfortuantely, no device is 100\% free of errors
in this transformation from quantum state to bitstring output. Next we describe
a simple model to cope with this classical readout channel. Although
there have been other procedures which are
scalable~\cite{Kandala2017,Dumitrescu2018,Magnard2018}, here we exploit our
small number of qubits to make the least number of assumptions on the channel
and perform an exponential number of experiments, which need to be done once to
characterize the channel.

The simple {\it error readout correction} implemented here consist of the
measurement characterization of all the $2^N$ projectors $\Pi_x = |x \rangle
\langle x |$, with $x \in \{0,1 \}^4$ for the (2,2)BAS realization here. Suppose
you trivially prepare any of the $2^4$ states of the computational basis, e.g.,
$x=0010$. In a noiseless scenario that measurement should yield a bitstring $x =
0010$ in the classical register for every single shot, allowing the computation
of the $P_x = 1.0$, as expected from preparation. However, in the experiment the
readouts correspond to different elements $y \in \{0,1 \}^4$ with a distribution
$p(y|x)$ due to assignment errors. In our procedure, we use a large number of
readouts (set to 10000 shots) to compute the distribution $p(y|x)$, after
trivial preparation of each $x \in \{0,1 \}^4$ .  Each of the $p(y|x)$ is used
as the $x$-th column of the transition matrix $M$. Thus, any quantum state with
population $P_x$ is related with actual observed output distribution $P_y$ as
$P_y = M P_x$. To recover the original distribution from the quantum state, we
invert the relation between $P_x$ and $P_y$, i.e. $P_x = M^{-1} P_y$. Before
each batch of five of learning curves, we calculate the inverse
transformation $M^{-1}$ and apply this to the readout distribution; this defines
a post-measurement process before the score and optimization steps. 

\begin{figure}[H]  
\centering 
  \includegraphics[width=0.95\textwidth]{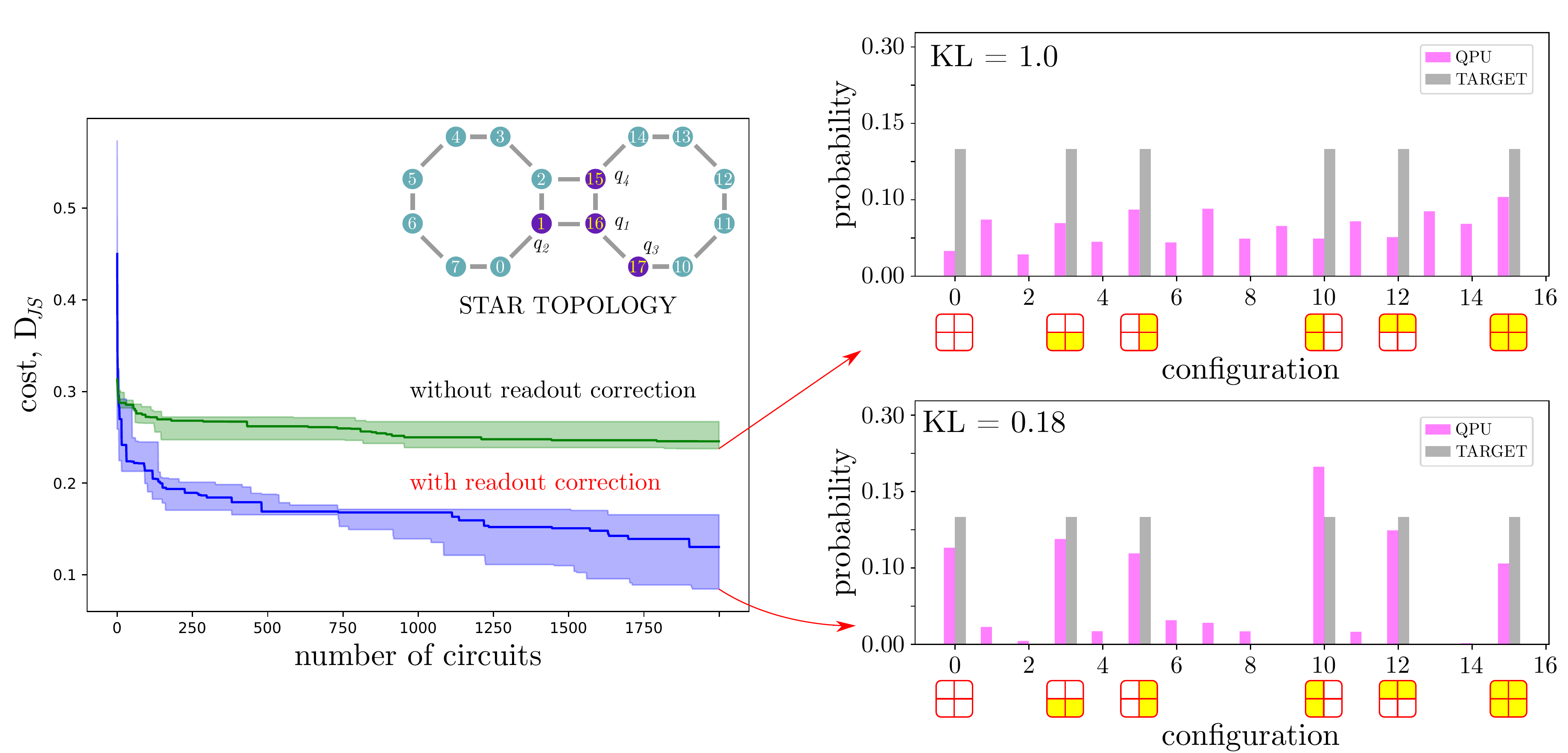} 
  \caption{{\it Comparison of DDQCL with and without readout
  correction.} Full learning curve is shown (left) with and without readout
  correction (blue and green line, respectively). Without error correction, the
  minimal value reached for the KL divergence is $\sim 1.0$, which is closer to
  a random distribution, and far from the target one (see histograms on the
  upper right). On the other hand, the training curve with readout correction
  reached a value of $KL \sim 0.13$, which is a significant improvement in the
  DDQCL.}  
    \label{fig:finet}
\end{figure}

To test the efficacy of this procedure, in Figure \ref{fig:finet}, we present
the training with the bare and the readout corrected training. The latter shows
significant improvement with a minimal score of KL $\sim 0.13$ from $\sim 1.0$
obtained using bare readouts. Therefore, we adopted the readout correction for
all the experiments reported in the main text.

\section{In-silico simulations}\label{s:simulations}

\begin{figure}[H]  
  \centering 
  \includegraphics[width=0.95\textwidth]{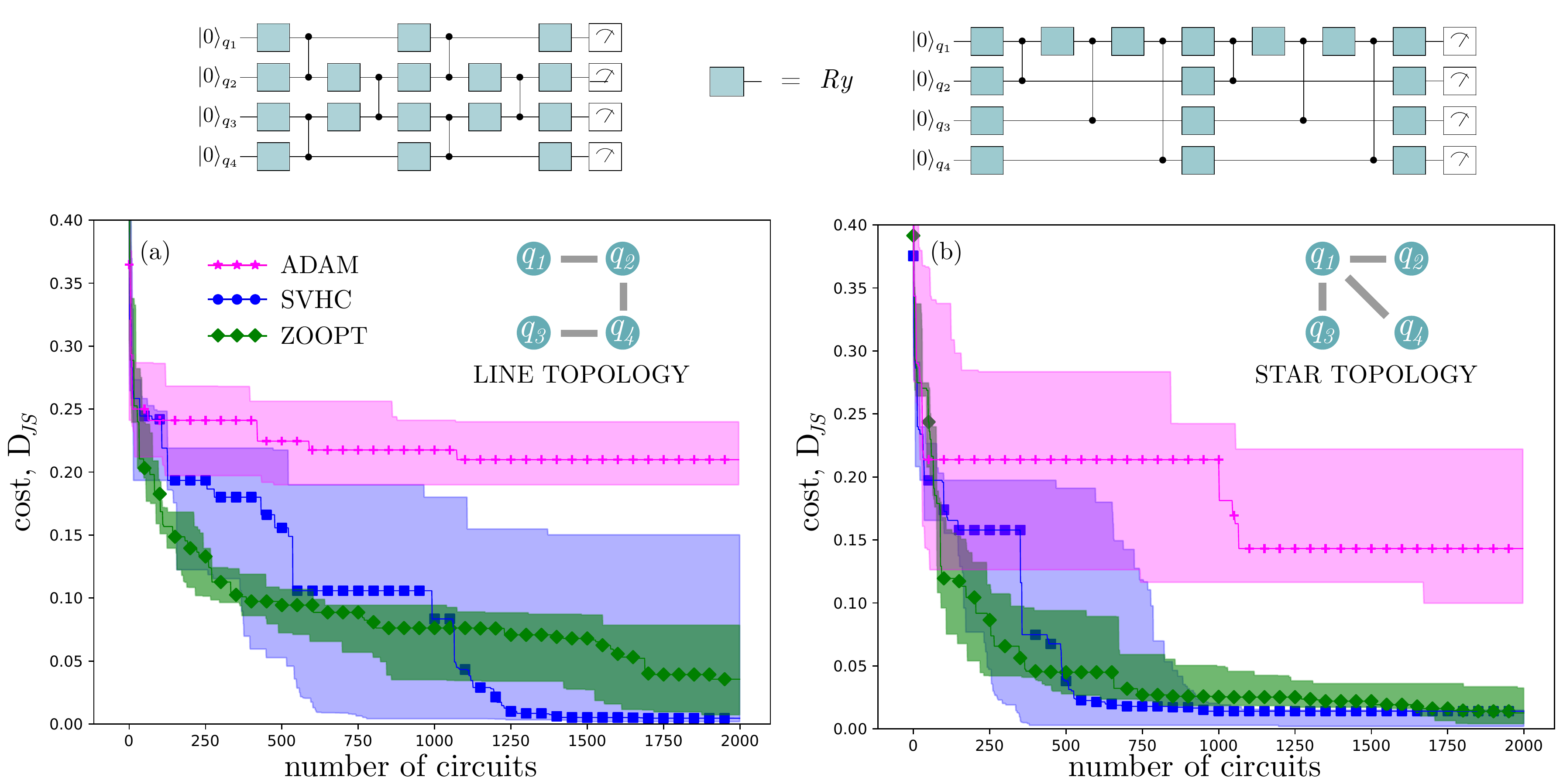}
  \caption{DDQCL simulations \textit{in-silico}, and under the assumption of
    noiseless qubits, but taking into account the stochasticity from finite
    readouts. In this simulations we used 3000 shots per circuit.}   
    \label{f:sim} 
\end{figure}

\begin{table}[h] 
\caption{Performance figures of merit from {\it in-silico} simulations of the
different HQC features explored in this work.}
\label{t:simulations} 
\includegraphics[width=0.5\textwidth]{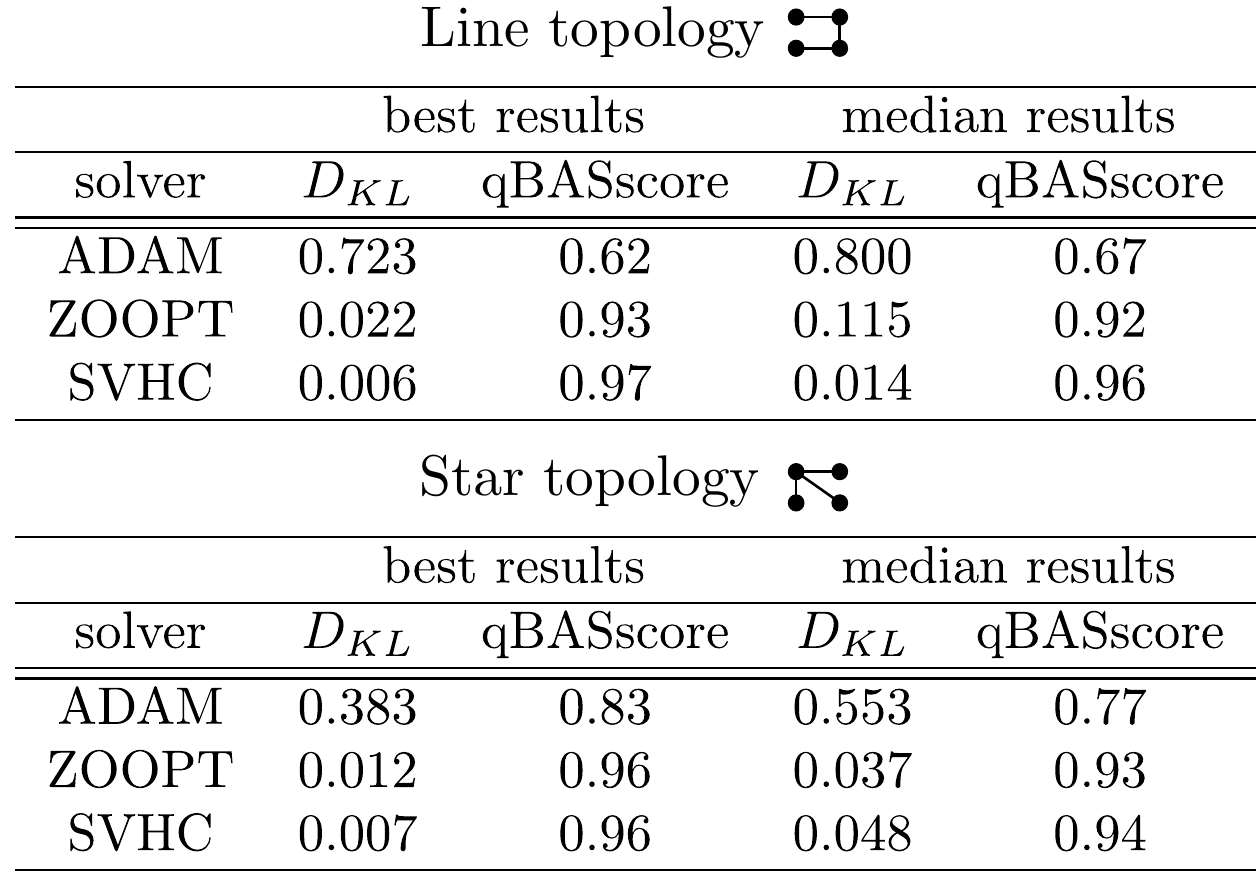}
\end{table}

\end{document}